\documentclass[12pt]{article}
\def\Journal#1#2#3#4{{#1} {\bf #2}, #3 (#4)}

\def\HS{{\cal H}}
\let\H\HS
\def\op#1{\mbox{\boldmath $#1$}}
\def\Tr{\rm Tr}
\def\TE{{\cal U}}
\def\dt{\partial_t}
\def\Ref#1{(\ref{#1})}
\def\G{\cal G}
\def\x{\vec{x}}
\def\pp{\psi^{\,\prime}}
\def\F{{\cal F}}
\let\ds\displaystyle
\def\J{{\bf J}}

\title{Remarks on a Nonlinear Quantum Theory}

\author{H.-D. Doebner \\[5pt]
Arnold-Sommerfeld Institute for Mathematical Physics,\\ Technical
University of Clausthal,\\ D-38678 Clausthal-Zellerfeld, Germany, \\e-mail: 
{\tt ashdd@pt.tu-clausthal.de}}

\date{March 2, 1998}

\begin{document}

\maketitle

Quantum mechanics in an intrinsically linear theory.
Consider a system localized and moving 
(non-relativistically) on a configuration space $M$.
Its quantization is based on the following building
blocks which are modeled on a separable Hilbert space $\HS =
L^2(M,d\mu)$ with $t$-dependent elements $\psi_t$: 
\begin{enumerate}
\item[(1)]
  {\bf Observables} are given by self-adjoint operators $\op{A}$. 
  The spectral representation of $\op{A}$
  allows a probability interpretation. 
\item[(2)]
  {\bf States} are described through 
  positive trace class operators (density matrices)
  $\op{W}_t$ ($\Tr\op{W}_t=1$) on
  $\HS$ which can be non-uniquely decomposed into projection operators 
  $\op{P}_{\psi_{j,t}}$ onto 1-dimensional subspaces (pure states)
  \begin{equation}
  \label{Wdec}
    \op{W}_t = \sum_{j\in I} \lambda_j \op{P}_{\psi_{j,t}}\,,~~ 
    \sum_{j\in I} \lambda_j = 1, ~ \lambda_j > 0.
  \end{equation}
  The set $\{\lambda_j,\psi_{j,t}\}_{j\in I}$ is denoted as a 
  {\em mixed state}; they are {\em physically equivalent}, if they 
  correspond to 
  the same $\op{W}_t$. Hence, $\op{W}_t$ defines an 
  equivalence class $C_t$
  in the set of mixed states.
\item[(3)]
  One {\em wants} the {\bf time dependence} of $\op{W}_t$ to be 
  fixed by the time
  dependence of pure states $\op{W}_{\psi_t}$, i.e., 
  through a smooth
  operator $\TE_{t_2,t_1}$ 
  (not necessarily linear)
  $ \TE_{t_2,t_1}\left(\psi_{t_1}\right) = \psi_{t_2}$. 
  With probability conservation
  $
    \|\psi_{t_2}\| = \|\TE_{t_2,t_1}\left(\psi_{t_1}\right)\| =
    \|\psi_{t_1}\| 
  $ and 
  time translation invariance one has
  $ \TE_{t}:= \TE_{t,0}$ and an evolution equation
  $ i\hbar\dt \psi_t = H(\psi_t)$, where 
  $ H =
  i\hbar \partial_t \TE_{t}$
  is a (not necessarily linear) 'Hamiltonian'.
  Because pure states rule the time evolution
  of $\op{W}_t$ only those $\TE_t$ 
  are allowed, which
  yield
  a unique evolution of $\op{W}_t$, i.e., the $\TE_t$ evolve
  an equivalence
  class $C_{t_1}$ into another equivalence class $C_{t_2}$, $t_2 > t_1$. 
  An evolution of this special equivalence classes is 
  in general only possible {\bf if}  $\TE_{t}$ yields a linear evolution
  equation; for exceptions see \cite{z1}.
  Non-linear $\TE_t$ yield {\bf acausal} effects \cite{z2}.
\item[(4)]
 One wants the systems to contain $N$ {\bf 1-particle
 systems}
 as {\bf building blocks}. 
 Realize $\H$ as a product of 1-particle spaces $\H^1$ and assume that
 the time evolution $\TE^{(N)}$ acts on product states as
 \begin{equation}
 \label{prod-evol}
 \TE^{(N)} [\psi_1 \otimes \cdots \psi_N] ~ = ~ \TE^{(1)} [\psi_1] \otimes 
 \cdots \otimes \TE^{(1)} [\psi_N] 
 \end{equation} 
 This assures, {\bf if} $\TE^{(1)} $ is linear, that in the absence of
 interparticle
 interaction terms initially uncorrelated subsystems remain
 uncorrelated and that $\TE^{(N)}$ can 
 consistently
 be extended (by linearity) to $\H$.
 For a non-linear evolution $\TE^{(1)}$, one has to define $\TE^{(N)}$
 on non-product states 
 through additional assumptions or -- equivalently -- to construct
 a hierarchy for the evolution of $1,\dots,N$ particles, i.e., for 
 $\TE^{(i)}$ with $i=1,\dots,N$.
 In general {\bf non-local effects} may appear.\cite{luckpp}
\end{enumerate}

The discussion shows that a nonlinear extension (NLE) of the linear
evolution (LE) yields in general difficulties:
\begin{itemize}
\item A non-linear 
$\TE^{(1)}$ is inconsistent with  density matrices as a model of mixed states.
\item N-particle systems built from 1-particle systems with non-linear
evolution are not fully separable.
\item Observables related to time
evolution may be realized through
non-linear operators for which a probability interpretation is not ad hand.
\end{itemize} 

One needs new concepts for {\bf a framework for a non-linear quantum theory},
e.g., 
for mixed states\cite{miel},
for the hierarchy of N-particle evolution equations\cite{z3} 
and a method how to interpret 
non-selfadjoint
operators.
A motivation to develop such a framework is that {\em ``The possibility of a
future non linear character of the quantum mechanics must be admitted,
of course''}
\cite{z4} and the observation that {\em ``all linear equations
describing the 
evolution of physical systems are known to be approximations of some 
nonlinear 
theories, with only one notable exception of the Schr\"odinger equation ``}
\cite{z5}. Furthermore a
convincing nonlinear extension of a linear time evolution is needed, which
is based on first principles. 

Such a -- more fundamental -- approach was 
given in\cite{z6}. The authors observe first that
all actual measurements are obtained from
positional measurements made at different times, i.e., the system is 
fully described through
the positional probability density
$\rho_t\,(m) = \overline{\psi_t(m)}\psi_t(m)$,
for all $m \in M$ and all $t$.
They are interested secondly in transformations $N$ of $\psi_t$ 
which
leave $\rho_t\,(m)$
invariant
\begin{equation}
\label{trafos}
 \overline{N[\psi_t](m)}N[\psi_t](m)=
 \overline{\psi_t(m)}\psi_t(m)\,. 
\end{equation}
and they argue that a system with states $\psi_t$ obeying
a (linear) evolution equation, and one with
states $N[\psi_t]$ obeying a transformed
equation, have the same physical content.
$N$ can be nonlinear and
nonlinear $N$ will
transform a system with LE 
into a physically equivalent system with
NLE which is, of course, {\em linearizable}.

To construct such  linear and nonlinear $N[\psi]$
we have from \Ref{trafos}
$$
  N[\psi_t](m) = \exp [i G_{\psi_t} (m)] \psi_t(m).
$$
We assume that 
$G$ is {\bf local} in the sense that
$ N[\psi_t](m) = N_F [\psi_t](m) \equiv \exp [i
F(\psi_t(m), m, t)] |\psi_t(m)|$ and that
$N_F$ fulfills the {\bf separation condition} $ N^{(N)}_{F}[\psi] =
    N_{F}[\psi_1] \otimes\cdots\otimes
    N_{F}[\psi_N]\,$.
Then one can prove ($|\psi|=R, {\rm arg} \psi = S$)
\begin{equation}
\label{nl-gauge}
  N_F[\psi] ~ = ~
  N_{(\gamma,\Lambda,\theta)}[\psi] = R
  \exp\left[i\left(\gamma(t)\ln R +\Lambda(t)S
      +\theta(m,t)\right)\right]\,.  
\end{equation}
The $N_{(\gamma,\Lambda,\theta)}$ 
are called {\bf non-linear gauge transformations} and 
form a group $\G$.
With these $N$ we construct in $R_x^3$ from 
$i\dt\psi_t = \left(\nu_1 \Delta + \mu_0 V\right)\psi_t$
a nonlinear -- but
linearizable -- evolution equation for
$ \pp(\x,t) = N_{(\gamma,\Lambda,\theta)}[\psi_t](\x)$.
From \Ref{nl-gauge} we get $(\theta \equiv 0)$
a family $\overline{\F_0}$ of NLEs which is a special case of \Ref{NLSE-2}
with 6 coefficients which are not independent but
constrained, depending on  $\nu_1,\mu_0$, and on
$\Lambda(t),\gamma(t)$.

To get a generic non-linearizable NLE
break the constraints and get a family ${\cal F}_1$; 
close this family in respect to the nonlinear
gauge transformations $\G$ and get $\overline{\F_1}$ again with constrained
coefficients.
Break the constraints, close and get $\overline{\F_2}$.
Continue with this process (called gauge generalization). 
After four steps we find a 10-parameter family with independent coefficients
and which is invariant under $\G$:
\begin{equation}
\label{NLSE-2}
\begin{array}{rcl}
  i \dt\pp&=&\ds \left(\nu_1\Delta + \mu_0 V\right)\pp 
            + i\nu_2R_2[\pp]\,\pp + \sum_{i=1}^5 \mu_i
              R_i[\pp] \\
  &&     + \alpha_1 \log|\pp|^2 \pp 
            + \alpha_2 (\arg\pp) \pp\,. 
  \end{array}
\end{equation}
with
\begin{equation}
\label{Rj}
\begin{array}{c}
\ds  \rho \, R_1[\psi] = \nabla\cdot\J\,,  \quad
 \rho \,  R_2[\psi] = \Delta \rho\, , \quad 
\rho^2 \,   R_3[\psi] = \J^2\,,\quad \\[5pt]
\rho^2  \, R_4[\psi] = \J\cdot\nabla\rho\,,\quad
\rho^2\,   R_5[\psi] = (\nabla\rho)^2\,, 
\end{array}
\end{equation}
with $\J$ the usual 
probability current.
The terms $R_i, ~ i=1,\dots,5$ were derived in a mathematically
and physically different approach in\cite{z8}, 
$\alpha_1 \log |\psi|^2 $ is the ansatz of \cite{z5} and  
$\alpha_2 (\arg\psi)$ of \cite{kostin}.

The last results and arguments suggest that nonlinear evolution
equations for pure states can be motivated from the observation that
all actual physical measurements are measurements of position and
time; this observation yields non-linear gauge transformations, which
applied to the linear Schr\"odinger equation leads via gauge
generalizations to a family of nonlinear evolutions with interesting 
mathematical and physical properties.

{\em It seems that there is a motivation for nonlinear quantum mechanics
and a path for its formulation. However, the path is not unique and -
up to now - there is no experimental evidence for a nonlinear
formulation of quantum mechanics.}


\end{document}